\documentstyle[aas2pp4,epsfig,12pt]{article}

\def\x0614{4U\,0614+09}
\def\mm{M\'endez }

\begin{document}

\title{Relations Between Timing Features and Colors in the 
X-Ray Binary 4U 0614+09}

\author{Steve van Straaten\altaffilmark{1}, Eric C. Ford\altaffilmark{1}, Michiel van der Klis\altaffilmark{1}, Mariano \mm\altaffilmark{1,2}, Philip Kaaret\altaffilmark{3}}

\authoremail{straaten@astro.uva.nl}
\altaffiltext{1}{Astronomical Institute, ``Anton Pannekoek'',
University of Amsterdam, Kruislaan 403, 1098 SJ Amsterdam,
The Netherlands.}
\altaffiltext{2}{Facultad de Ciencias Astron\'omicas y Geof\'{\i}sicas,
Universidad Nacional de La Plata, Paseo del Bosque S/N, 1900 La Plata,
Argentina.}
\altaffiltext{3}{Harvard--Smithsonian Center for Astrophysics, 
60 Garden Street, Cambridge, MA 02138, USA.}

%\centerline{\bf {XXXXXXXXXXXXX}}

\begin{abstract}

We study the correlations between timing and X-ray spectral properties
in the low mass X-ray binary 4U 0614+09 using a large (265-ks) data
set obtained with the Rossi X-ray Timing Explorer.  We find strong
quasi-periodic oscillations (QPOs) of the X-ray flux, like the
kilohertz QPOs in many other X-ray binaries with accreting neutron
stars, with frequencies ranging from 1329 Hz down to 418 Hz and,
perhaps, as low as 153 Hz.  We report the highest frequency QPO yet
from any low mass X-ray binary at 1329 $\pm$ 4 Hz, which has
implications for neutron star structure. This QPO has a 3.5 $\sigma$
single-trial significance, for an estimated 40 trials the significance
is 2.4 $\sigma$.  Besides the kilohertz QPOs, the Fourier power
spectra show four additional components: high frequency noise (HFN),
described by a broken power--law with a break frequency between 0.7
and 45 Hz, very low frequency noise (VLFN), which is fitted as a
power--law below 1 Hz, and two broad Lorentzians with centroid
frequencies varying from 6 to 38 Hz and 97 to 158 Hz, respectively. We
find strong correlations between the frequencies of the kilohertz
QPOs, the frequency of the 6$-$38 Hz broad Lorentzian, the break
frequency of the HFN, the strength of both the HFN and the VLFN and
the position of the source in the hard X-ray color vs. intensity
diagram.  The frequency of the 97$-$158 Hz Lorentzian does not
correlate with these parameters.  We also find that the relation
between power density and break frequency of the HFN is similar to
that established for black hole candidates in the low state. We
suggest that the changing mass accretion rate is responsible for the
correlated changes in all these parameters.

\end{abstract}


\section{Introduction} 

Accretion in neutron star low mass X-ray binaries can be studied
through the spectral and timing properties of the associated X-ray
emission. Some of these systems exhibit quasi-periodic oscillations
with frequencies ranging from a few hundred Hz to more than 1000 Hz
(kilohertz QPOs). These oscillations have been observed with Rossi
X-ray Timing Explorer (RXTE; for reviews and references see van der
Klis 1998, 1999; Swank 1998 and
http://www.astro.uva.nl/$\sim$ecford/ qpos.html). The high frequency of
these signals nearly certainly reflects the short dynamical time scale
in the region near the compact object where they originate. The timing
properties at low frequencies ($\nu < 100$ Hz) as well as the spectral
properties are the basis of a classification of the systems as {\em Z}
or {\em Atoll} sources (Hasinger \& van der Klis 1989). The low
frequency part of the Fourier power spectra of the atoll sources is
dominated by two features (first identified with the EXOSAT
satellite): a power--law red noise component in the lowest range of
the spectrum ($\nu <$ 1 Hz) called very low frequency noise (VLFN),
and a high frequency noise (HFN) component which is flat at low
frequency and breaks to follow a power--law with an index of about 1
at higher frequency. In some cases, a low frequency Lorentzian is also
seen (e.g. Wijnands \& van der Klis 1999a). The HFN is very similar in
shape to the low state noise of the black hole candidates (BHCs; van
der Klis 1994a; Yoshida et al. 1993). Also the correlation between the
break frequency of the HFN and the centroid frequency of the low
frequency Lorentzian is the same for black hole candidates and neutron
stars (Wijnands \& van der Klis 1999a).

These timing properties are connected with the properties of the X-ray
energy spectrum. This is seen very clearly in the Z sources both for
the lower frequency features (e.g., van der Klis et al. 1985; Hasinger
\& van der Klis 1989; Wijnands et al. 1997a; Kuulkers et al. 1997;
Kuulkers et al. 1996) and for the kilohertz QPOs (van der Klis et
al. 1996; Wijnands et al. 1997b; Jonker et al. 1998).  Similar
correlations exist for atoll sources as well, both for the lower
frequency features (e.g., Hasinger \& van der Klis 1989; Prins \& van
der Klis 1997; \mm et al. 1997) and for the kilohertz QPOs (e.g., Ford
et al. 1997b; Kaaret et al. 1998; \mm et al. 1999). Also connections
have been made between the frequency of the kilohertz QPOs and the
frequencies of the lower frequency features (van der Klis et al. 1996;
Ford \& van der Klis 1998; Wijnands \& van der Klis 1998). The most
obvious reason for all these correlations is that the changing mass
accretion rate causes changes in the timing and spectral properties
(Hasinger \& van der Klis 1989).

4U 0614+09 has been studied previously with EXOSAT (Singh \& Apparao
1994; Barret \& Grindlay 1995).  The EXOSAT data showed an
anti-correlation of high and low energy emission (Barret \& Grindlay
1995), and observation with the Burst and Transient Source Experiment
(BATSE) and the RXTE all--sky monitor showed that this
anti-correlation extends to about 100 keV (Ford et al. 1996). X--ray
bursts from 4U 0614+09 were identified by OSO-8 (Swank et al. 1978)
and WATCH (Brandt et al. 1992; Brandt 1994). The only previously
published study investigating the broadband ($\sim$0.03--1200 Hz)
X--ray power spectra is the \mm et al. (1997) study of 14 ks of RXTE
data, concluding that this source can be classified as an atoll
source. Strong kilohertz QPOs were discovered in this source with RXTE
(Ford et al. 1997a). The frequency of these QPOs correlated with count
rate over short time scales, but on long time scales this correlation
broke down (Ford et al. 1997a; \mm et al. 1997). A more robust
correlation was shown to exist between the frequency and either the
flux of a spectral component, the X-ray spectral shape, or a hardness
ratio (Ford et al. 1997b; Kaaret et al. 1998; \mm et al. 1997). The
physical implication is that both these latter parameters and the
frequency of the kilohertz QPOs are determined by a changing mass
accretion rate. The 265 ks of RXTE data we use here includes 75 ks
previously used by Ford et al.  (1997a;1997b) and Kaaret et al. (1998)
and 14 ks by \mm et al. (1997).

In this paper we show a connection between the changes in spectral
shape and both the kilohertz QPOs and the lower frequency features in
a large RXTE data set for 4U 0614+09. In \S 2 we describe our analysis
of the spectral and timing data. In \S 3 we present our results and in
\S 4 we compare our results to other atoll sources and discuss the
connections to physical models.

\section{Observations and Data Analysis}

In this analysis we use data from RXTE's proportional counter array
(PCA; for more instrument information see Zhang et al. 1993).  The
total observing time is about 265 ks, approximately 75\% of the total
RXTE observing time on 4U 0614+09 to date. Data sets are split into
near continuous time intervals of approximately 2500 s which we shall
call observations. These observations are listed in Table 1.  We use
the 16--s time--resolution Standard 2 mode to calculate the colors (as
defined below) and the 122$-\mu$s time--resolution Event and
0.95$-\mu$s time--resolution Good Xenon modes for Fourier analysis. No
X--ray bursts were observed in this data set.

\begin{figure*}
\figurenum{1}
{\psfig{figure=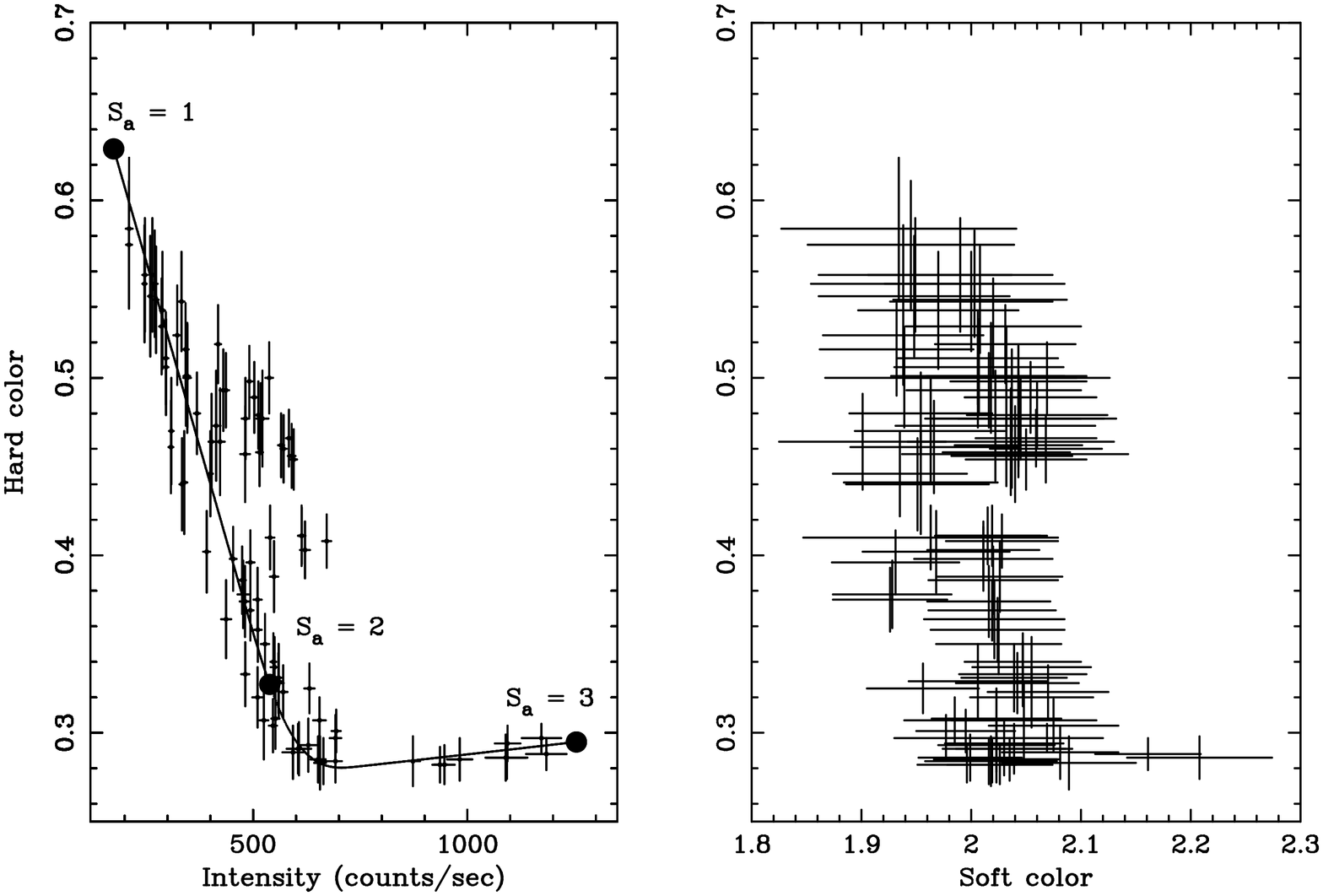,angle=270,width=6.5in}}
\caption{Hard color vs. intensity (left panel) and soft color (right panel) for 4U 
0614+09. Hard color is defined as the ratio of count rates in the energy bands 
9.7$-$16.0$/$6.0$-$9.7 keV. Soft color is defined as the ratio of count rates in 
the energy bands 3.5$-$6.0$/$2.0$-$3.5 keV. Intensity is 
defined as the count rate in the energy band 2.0$-$16.0 keV. Each point is an average
over one observation (typically 2500 s). Circles in the left panel indicate values 
of $S_{\rm a}$ along the spline curve (see text).}
\label{fig:hidcd}
\end{figure*}

We calculate a hard color, defined as the count rate in the energy
band 9.7$-$16.0 keV divided by the rate in the energy band 6.0$-$9.7
keV, and a soft color, defined as the count rate in the energy band
3.5$-$6.0 keV divided by the rate in the energy band 2.0$-$3.5 keV.
We also calculate the intensity, the count rate over five detectors in
the energy band 2.0$-$16.0 keV. To obtain the count rates in these
exact energy ranges we interpolate linearly between count rates in the
PCA channels. We calculate the colors and intensity first for each
time interval of 16 s. We subtract the background contribution in each
band using the standard background model for the PCA version 2.1b. We
then calculate the average and standard deviation for each observation
among the 16 s points.  In the left panel of Figure \ref{fig:hidcd} we
show a diagram of hard color vs. intensity, in the right panel a
diagram of hard color vs. soft color. Each point in the plot
corresponds to the average of the 16 s points in one observation. The
error bars indicate the standard deviation from the 16 s intervals. In
the few observations where only three or four of the five PCUs were
on, we normalize the count rates to five detectors.

Most of the data fall on one curve. However, there is a set of points
in the $0.4-0.5$ hardness range that appears to be shifted to higher
count rate.  These are the points from the 1997 observations. There is
no obvious reason for the shift.  The general trend in the soft color
vs. hard color diagram is the same as in the diagram of hard color
vs. intensity including the offset of the 1997 points.  We use the
diagram of hard color vs. intensity in this source instead of the
color--color diagram, as was done with 4U 1608$-$52 (\mm et al. 1999)
and 4U 1728$-$34 (\mm \& van der Klis 1999), because there is less
scatter in count rate than in soft color.

Most of our observations occurred in gain {\em epoch 3} (between 1996
April 15 and 1999 March 22).  Four of our observations (a total of 9.4
ks of data) took place in gain {\em epoch 1} (before 1996 March 21),
two (5 ks) in gain {\em epoch 2} (1996 March 21 -- 1996 April 15) and
two (6.9 ks) took place in gain {\em epoch 4} (1999 March 22 to
present).  In order to correct the colors for these data for the
changes in instrumental response, we calculate colors of the Crab
nebula, which can be supposed to be constant in its colors, using the
same energy bands in the various gain epochs and calculate a
correction factor which we apply to the data in epochs 1, 2, and 4 to
make the Crab colors match those of epoch 3 (c.f. Kuulkers et
al. 1994).  The correction factor for hard colors for epochs 1,2 and 4
is about 0.2 \%, 1.0 \% and 0.2 \% respectively.  The correction
factor for soft colors for epochs 1,2 and 4 is about 3.5 \%, 1.0 \%
and 20 \% respectively.  Finally, the correction factor for
intensities for epochs 1,2 and 4 is about 1.0 \%, 6.0 \% and 2.0 \%
respectively. We have tested the validity of these corrections for
epochs 1 and 2 by using spectral fits of 4U 0614+09 observations and
the response matrices to calculate cross-epoch corrections. The
correction factors obtained in this way are within 1.5 \% larger in
hard color, within 7 \% larger in soft color and within 10 \% smaller
in intensity as the Crab-based corrections.

We parametrize the position of each observation in the hard color
vs. intensity diagram by a variable, $S_{\rm a}$, which measures the
distance along the curve traced out by the data (see Figure
\ref{fig:hidcd}). This is inspired by the work on Z sources where the
distance along the Z track in the color--color diagram is
parameterized by $S_{\rm z}$ (Hasinger et al. 1990; Hertz et al. 1992)
and by recent results on the atoll sources 4U 1608$-$52 (\mm et
al. 1999) and 4U 1728$-$34 (\mm \& van der Klis 1999). We note that
these previous studies use color--color diagrams as opposed to the
color--intensity diagram we use here.

In calculating $S_{\rm a}$, we choose a few points in the
hardness--intensity diagram and fit a quintic spline through these
points and define $S_{\rm a}$ = 1 as the point (174,0.629) and $S_{\rm
a}$ = 3 as the point (1255,0.295) as indicated in the figure. We
rescaled the axes so that both axes range over the same values.  We
then find the $S_{\rm a}$ value of each observation by determining the
closest point on the curve.  The error in $S_{\rm a}$ is obtained by
quadratically adding the errors in $S_{\rm a}$ that resulted from
mapping the one sigma errors in hard color and intensity on to $S_{\rm
a}$.
 
\begin{figure*}
\figurenum{2}
% need rescaling in preprint mode
\epsscale{1.5}
\plotone{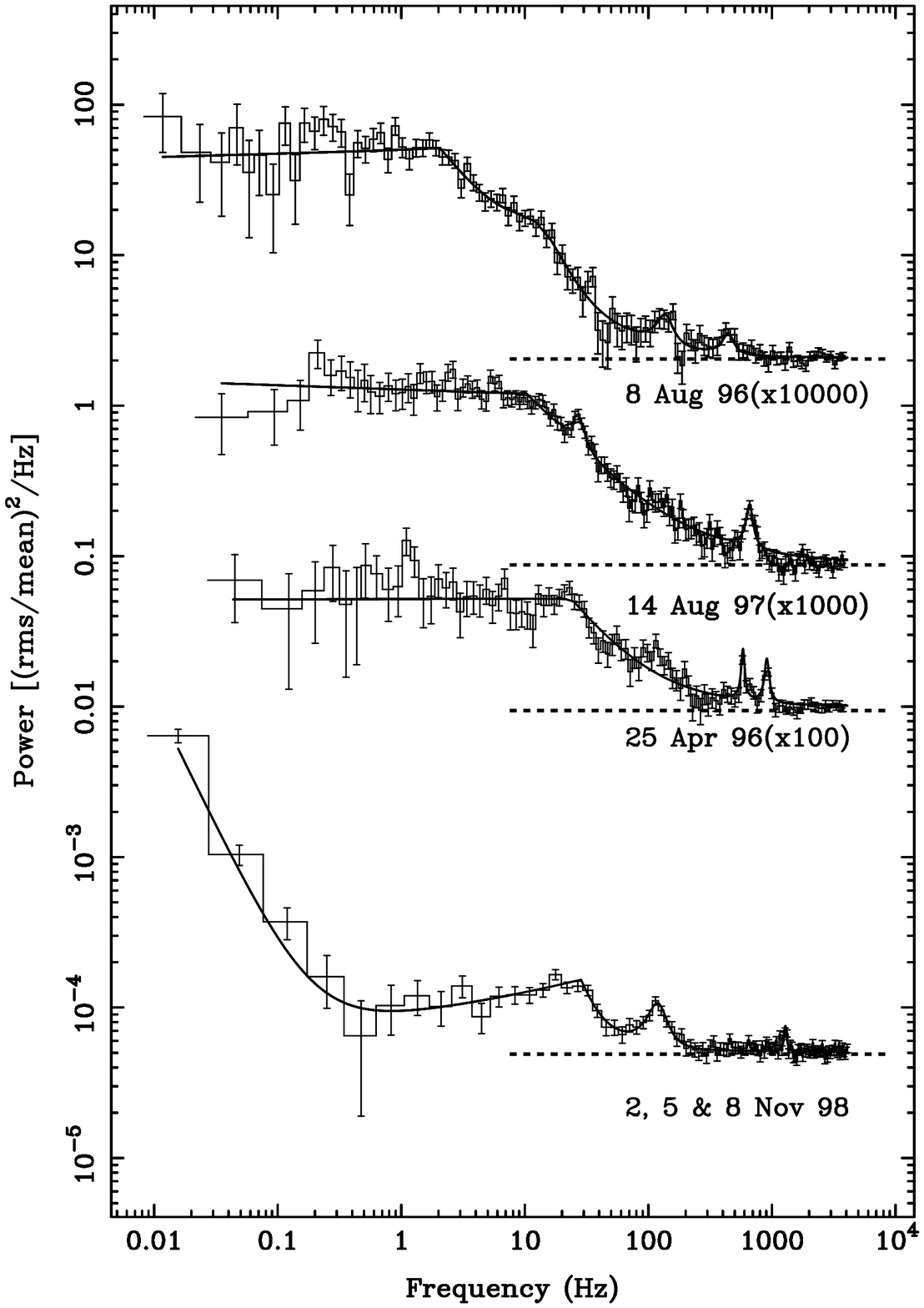} 
\caption{Representative power density spectra of 4U 0614+09 for four 
observations plotted from top to bottom in order of increasing $S_{\rm a}$ value, 
$S_{\rm a}$ is respectively 1.25, 1.71, 2.01 and 2.59. The power spectra are for the full 
energy band of the PCA. We subtract a constant value, somewhat less than the Poisson level and 
renormalize each spectrum to get fractional squared rms per Hz. The Poisson level is indicated 
for each spectrum with a dashed line. For clarity each spectrum is multiplied by a 
constant as indicated. Fit functions are shown which consist of 
a power--law at low frequency, a broken power--law, a low frequency 
Lorentzian, a Lorentzian near 100 Hz and kilohertz QPOs. The 25 April 1996 spectral fit does not 
include a $\sim$100 Hz Lorentzian, since it was less than 3 $\sigma$ significant.}
\label{fig:reprps}
\end{figure*}

We calculated power spectra from the high time resolution data. For
each observation we make two spectra. One in which we use all the
energy channels of the PCA, for the fit at low frequency ($\nu \leq$
400 Hz), and one in which we apply channel cuts to select high
energies for the fit of the kilohertz QPOs.  The QPOs are generally
stronger at higher energies.  For the kilohertz QPO fits we use a
nominal energy range of 4.6$-$97.3 keV (unbinned PCA channels 13$-$249
for gain epoch 3), for the other epochs we use the channel range
closest to these energies. For gain epoch 1 we use channels 17$-$249
(4.2$-$68.0 keV), for gain epoch 2 channels 13$-$249 (3.5$-$75.6 keV)
and for gain epoch 4 we use channels 11$-$219 (4.6$-$97.5 keV). In all
cases, we divide the data into segments of 64 or 256 s and then bin
the data in time before Fourier transforming such that the Nyquist
frequency is always 4096 Hz. We then average the individual power
spectra over each observation. Four representative power spectra are
shown in Figure \ref{fig:reprps}.  We fit each spectrum with a
function that was found to describe all the power spectra well. The
four main components are as follows.  (I) The Poisson level. For the
Poisson level we fit a value constant in frequency. This is valid
since dead-time effects are very small in this weak source. This value
is approximately 2 as expected from Poisson noise with the
normalization of Leahy et al. (1983). We fit the Poisson level from
2000 to 4000 Hz, and then fix it and fit the low frequency power
spectra up to 400 Hz. When we fit the kilohertz QPOs we let the
Poisson level float. (II) A broken power--law whose power, $P$, is
defined as:
$$ P(\nu) = \left\{ \begin{array}{ll}
A\nu^{-\alpha_0} & (\nu < \nu_{\rm break}) \\
A\nu_{\rm break}^{\alpha_1-\alpha_0}\nu^{-\alpha_1} & (\nu > \nu_{\rm break})
\end{array} \right. $$
where A, $\nu_{\rm break}$, $\alpha_0$ and $\alpha_1$ are the
normalization factor, break frequency and the index below and above
the break. Following previous work we will refer to this component as
high frequency noise (HFN; Hasinger \& van der Klis 1989). (III) A
power--law below 1 Hz called very low frequency noise (VLFN; Hasinger
\& van der Klis 1989). (IV) The kilohertz QPOs, described by
Lorentzians with best-fit centroid frequencies from 153 to 1329 Hz. In
one power spectrum the centroid frequencies of the QPOs fall in the
same frequency range as the low frequency noise ($\nu <$ 400 Hz), in
this case we include the shape of the low frequency noise in our fit
of the kilohertz QPOs.

In some cases we add two additional components to the fit function : a
Lorentzian with a best-fit centroid frequency between 6 and 38 Hz
(FWHM 2$-$19 Hz), and another Lorentzian with a best-fit frequency
between 97 and 158 Hz (FWHM 22$-$144 Hz).  Other atoll sources show
similar components in their power spectra, as seen with GINGA in 4U
1608$-$52 (Yoshida et al. 1993) and with RXTE in 4U 1705$-$44 (Ford,
van der Klis \& Kaaret 1998), 4U 1728$-$34 (Ford \& van der Klis
1998), 4U 1820$-$30 (Zhang et al. 1998a), 4U 1608$-$52 (\mm et
al. 1999; Ford et al. 1999) and in other binaries as well (Wijnands \&
van der Klis 1999a; Psaltis et al. 1999). The $\chi^{2}$/dof of our
fits are typically 1.0 to 1.4 with 127 to 164 degrees of freedom.

\section{Results}

The broken power--law component of the power spectra is similar to
that in previous measurements in other atoll sources and black hole
candidates. The rms fractional amplitude (0.1$-$400 Hz) ranges from 3
to 30 \%. The average value of $\alpha_0$ is close to zero ($-0.02$),
and the average value of $\alpha_1$ is close to unity (0.92). There is
no evidence for a second break in this component as is seen in some
descriptions of the power spectra of the BHCs and the atoll source 4U
1608$-$52 (Yoshida et al. 1993). Above $S_{\rm a}$ values of 2, the
broken power--law has a peaked shape (Hasinger \& van der Klis 1989):
$\alpha_0$ is between $-0.2$ and $-1.6$ and $\alpha_1$ is between 1
and 2. These power spectra could also be fitted with a broken
power--law with $\alpha_0\sim0$ and $\alpha_1\sim1$ if we add an extra
Lorentzian with a centroid frequency near $\nu_{\rm break}$. The fits
reported here, however, do not include this Lorentzian as it is less
than 3$\sigma$ significant in most cases.

Historically other functions have been used to describe this HFN
component: a cut-off power--law described by $P \propto
\nu^{-\alpha}\exp^{-{\nu}/{\nu_c}}$ (Hasinger $\&$ van der Klis 1989),
or Lorentzians with a central frequency of zero Hz (Grove et al. 1994;
Olive et al. 1998). We have tried such functions, but the $F$-test
statistics show no preference for one of the functions. We employ the
broken power--law because it uses fewer free parameters (with the
other functions we often need to add an extra Lorentzian with a
frequency near $\nu_{\rm break}$). Fits of the correlations (see
below) also have better $\chi^{2}$/dof with the broken power--law
function.

\begin{figure*}
\figurenum{3}
% need rescaling in preprint mode
\epsscale{1.5}
\plotone{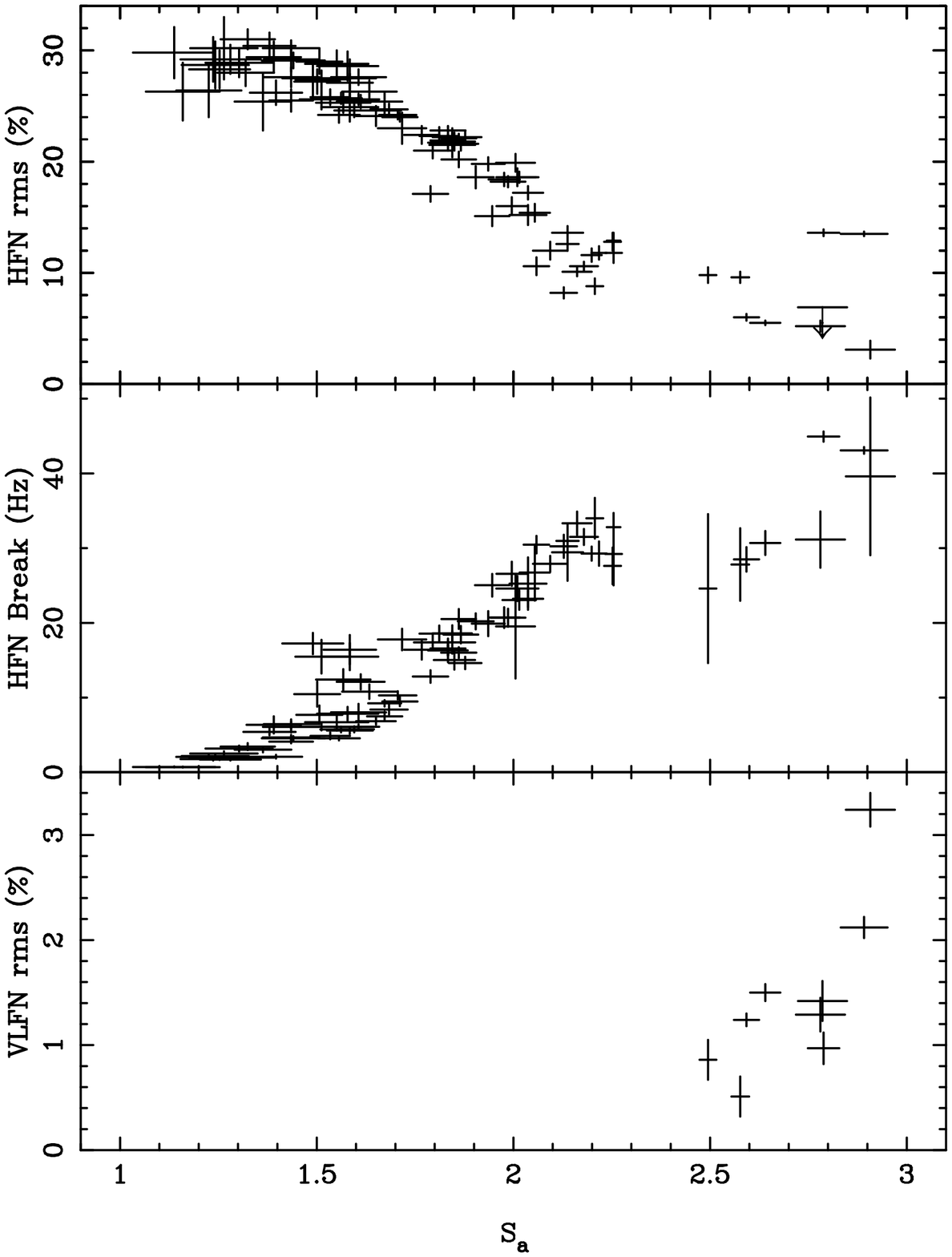} 
\caption{Parameters of the fit to the power spectra versus $S_{\rm a}$, 
the position in the hard color vs. intensity diagram (Fig. \ref{fig:hidcd}). The plots are 
rms fractional amplitude (0.1$-$400 Hz) of the high frequency noise (broken power--law), 
the frequency of the break in this component, and the rms fraction (0.01$-$1 Hz) 
of the very low frequency noise.}
\label{fig:salf}
\end{figure*}

The VLFN is present in some of the intervals of March, November and
December 1998 (see Table 2 and Fig. \ref{fig:reprps}) with rms
fractional amplitude (0.01$-$1 Hz) between 0.5 and 3.2 \%, when
$S_{\rm a} \geq 2.49$ (see Table 2 and Fig. \ref{fig:salf}).  This
state (extremely soft with VLFN) is often referred to as the {\em
banana} state (Hasinger \& van der Klis 1989). In all but one of the
observations where the VLFN was present, it is accompanied by a
significant HFN component with rms fractional amplitude between 3.1
and 13.6 \% (see Table 2).

\begin{figure*}
\figurenum{4}
% need rescaling in preprint mode
\epsscale{1.5}
\plotone{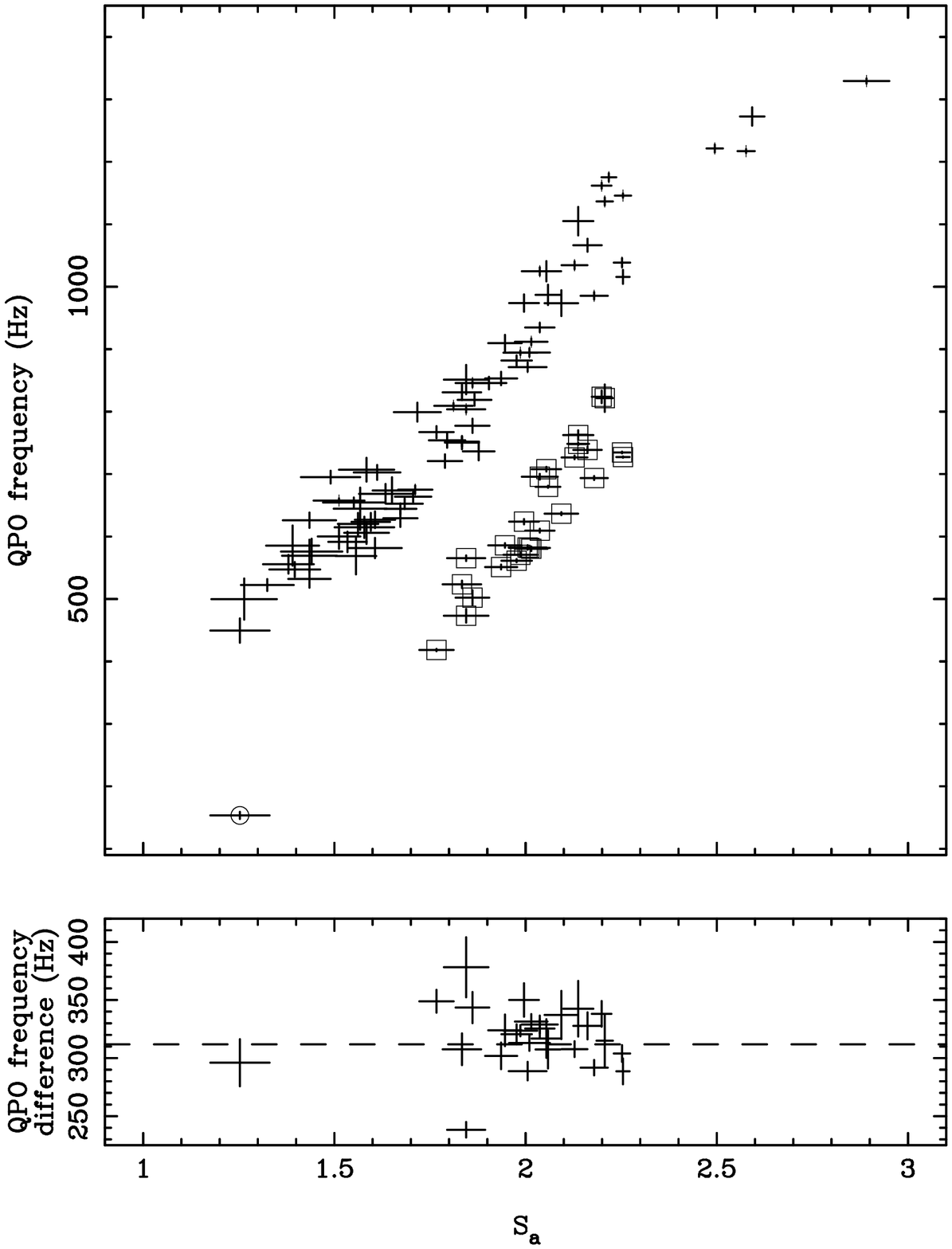} 
\caption{Frequency of the kilohertz QPOs vs. $S_{\rm a}$ (top) and difference in 
frequency between double QPOs vs. $S_{\rm a}$ (bottom). Plusses represent the
higher frequency peak, squares the lower one. The circle marks the case where the QPO 
can be identified as either the lower kilohertz QPO or the 97$-$158 Hz Lorentzian 
(see text and Fig. \ref{fig:saqpo}).}
\label{fig:sakhzqpo}
\end{figure*}

\begin{figure*}
\figurenum{5}
% need rescaling in preprint mode
\epsscale{1.5}
\plotone{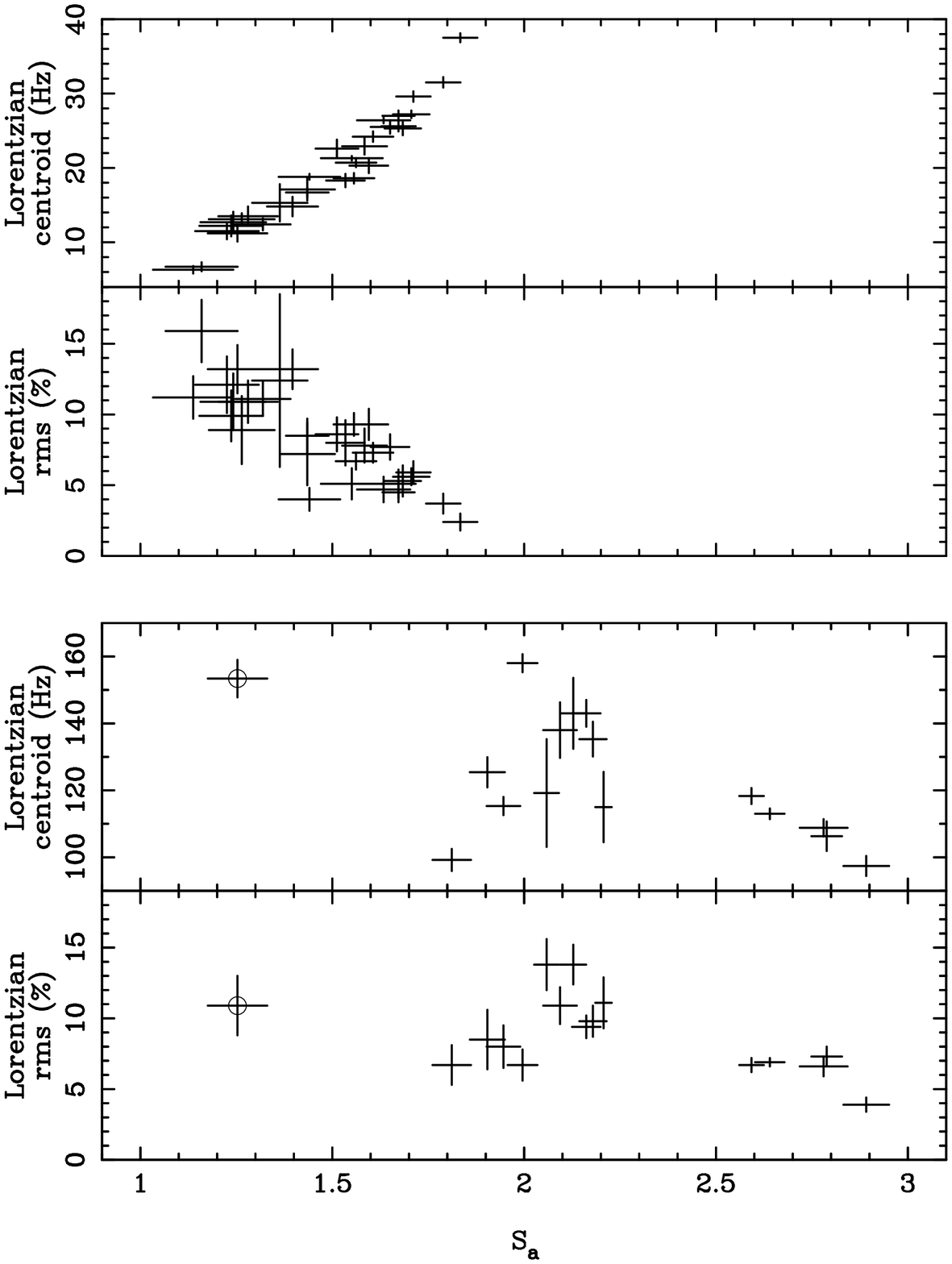} 
\caption{Centroid frequency and rms fractional amplitude of the 6$-$38 Hz (top) and 
97$-$158 Hz (bottom) Lorentzian vs. $S_{\rm a}$. The circle marks the case where the QPO can be identified 
as either the lower kilohertz QPO or the 97$-$158 Hz Lorentzian (see text and Fig. \ref{fig:sakhzqpo}).}
\label{fig:saqpo}
\end{figure*}

The kilohertz QPO parameters (fitted in the energy range 4.6--97.3
keV) are listed in Table 1.  The kilohertz QPOs have similar
properties to those in other low mass X-ray binaries.  The lower
frequency QPO ranges from 824 Hz down to 418 Hz and, perhaps, as low
as 153 Hz. The upper frequency QPO ranges from 1329 down to 449
Hz. Note that these are fits of the power spectra above 4.6 keV. We
include the kilohertz QPOs when the significance based on the error in
the integrated power is above 3.0 $\sigma$. Of the kilohertz QPOs
reported here, 69 \% have a significance of 3.5 to 8.9 $\sigma$ and 31
\% have a significance of 3.0 to 3.5 $\sigma$. The QPO with the
highest observed centroid frequency, 1329 $\pm$ 4 Hz is 3.5 $\sigma$
significant. This significance is for a single trial (van der Klis
1989). We estimate that the number of trials is about 40 based on the
width of the detection interval ($\sim$ 1000 Hz) and the width of the
QPO ($\sim$ 25 Hz) at 1329 Hz.  Including these trials reduces the
significance to 2.4 $\sigma$. We note that the QPO frequency of 1329
Hz deviates from the linear trend seen for the other kilohertz QPOs
(Fig. \ref{fig:sakhzqpo}).  The break frequency in the same
observation also deviates from the linear trend seen for the break
frequencies at $S_{\rm a}$ below 2.3 (Fig. \ref{fig:salf}).  The
frequency difference between the two kilohertz QPOs is constant over a
wide range in frequency in 4U 0614+09 and has a weighted average value
of 311.8 $\pm$ 1.8 Hz (Fig. \ref{fig:sakhzqpo}).  At least one QPO is
present over basically the entire range of $S_{\rm a}$ (1.25 to 2.89)
and in 74 of the 94 power spectra. At $S_{\rm a} \geq$ 2.49 a QPO is
present in 4 of 9 observations with rms fractional amplitudes ranging
from 5.1 to 11.1 \%.  There is generally only one QPO above $S_{\rm
a}$ of about 2.3 and below $S_{\rm a}$ of about 1.7.  There is one
exception at $S_{\rm a}$ = 1.25, where there is a QPO present at 153
$\pm$ 6 Hz. However this QPO can be identified either as the lower
kilohertz QPO or one of the 97$-$158 Hz Lorentzians (the circle point
in Fig. \ref{fig:sakhzqpo} and \ref{fig:saqpo}) on the basis of the
correlations of centroid frequency, rms fractional amplitude or FWHM
with $S_{\rm a}$.  In most cases where single QPOs are observed, they
can be identified as the upper (or, in one case lower) one on the
basis of the correlations of centroid frequency with $S_{\rm a}$
and/or hard color. The four highest frequency features are single QPOs
but are likely the higher frequency of the two QPOs based on the
$S_{\rm a}$ correlation. There is, however, some ambiguity in this
identification.

In 31 of the total of 94 observations we have found it necessary to
include a low frequency Lorentzian at 6 to 38 Hz. We include it in the
fit when the significance based on the error in its integrated power
is above 3.0 $\sigma$. This Lorentzian is seen only at low $S_{\rm a}$
and increases in frequency with $S_{\rm a}$
(Fig. \ref{fig:saqpo}). The rms fractional amplitude (2.4$-$15.9 \%)
decreases as the centroid frequency increases (Fig. \ref{fig:saqpo}).
In some cases the Lorentzian is broad, in the 1996 August 8 interval
for instance it has Q(=$\nu$/FWHM) of 1.3 $\pm$ 0.5.

The final component of our fit, a Lorentzian near 100 Hz, is clearly
present at a high $S_{\rm a}$ value (Fig. \ref{fig:saqpo}). We include
it in the fit when the significance based on the error in its
integrated power is above 3.0 $\sigma$ (15 observations). The best-fit
centroid frequencies are between 97 and 158 Hz and the rms fractional
amplitudes between 3.9 and 13.8 \%. Adding it to the fit makes
$\alpha_1$ and $\nu_{\rm break}$ of the HFN significantly larger.

Most of the features in the Fourier spectra discussed above are
correlated smoothly with the changing energy spectrum. This is
summarized in Figures \ref{fig:salf}, \ref{fig:saqpo} and
\ref{fig:sakhzqpo} where we plot several quantities vs.  $S_{\rm a}$,
which measures the changing colors. The same quantities do not
correlate smoothly with count rate (in the energy band 2.0$-$16.0
keV). The only exception to these correlations is the frequency of the
97$-$158 Hz Lorentzian, which does not correlate with $S_{\rm a}$
(Fig. \ref{fig:saqpo}).  The 97$-$158 Hz Lorentzians do not fit on an
extrapolation of the $S_{\rm a}$ vs. frequency correlation of the
6$-$38 Hz Lorentzians or its harmonics, suggesting that the 97$-$158
Lorentzian has a different physical origin.

\begin{figure*}
\figurenum{6}
% need rescaling in preprint mode
\epsscale{1.5}
\plotone{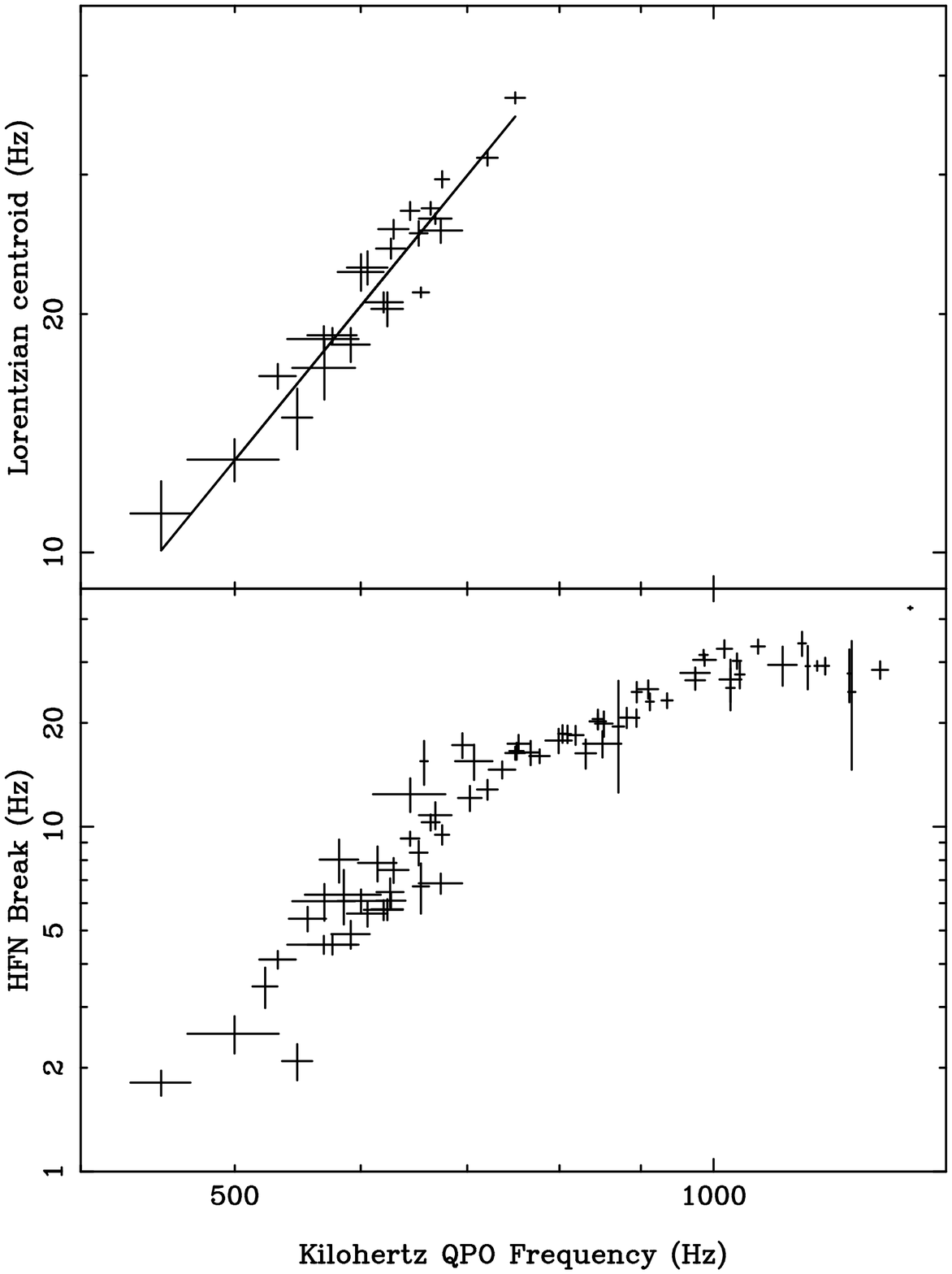} 
\caption{Frequency of the Lorentzian at low frequencies (top) and the break frequency of 
the broken power--law function (bottom) versus the
frequency of the upper kilohertz QPO. The function shown is a power--law
with index 2.46.}
\label{fig:khzvsqpo}
\end{figure*}

Since many of the parameters of the power spectra fits are correlated
with $S_{\rm a}$, many of these values are correlated with each
other. Two such correlations are shown in Figure
\ref{fig:khzvsqpo}. Both the centroid frequency of the 6$-$38 Hz
Lorentzian, $\nu_{\rm LFLor}$, and the break frequency of the HFN are
correlated with the centroid frequency of the higher frequency
kilohertz QPO, $\nu_{\rm kHz}$.  Fitting a relation $\nu_{\rm LFLor}$
= A$\nu^{\alpha}_{\rm kHz}$, we obtain $\alpha$ = 2.46 $\pm$ 0.07 and
A = 3.0($\pm 1.4) \times 10^{-6}$ (line in Fig. \ref{fig:khzvsqpo}).
The fit has a large $\chi^{2}$/dof of 11.6. The quoted errors use
$\Delta\chi^{2}$ = 1.0.

When we plot centroid frequency of the low frequency Lorentzian
vs. the break frequency, the points fall in the same region as the
plot that Wijnands \& van der Klis (1999a) established for the BHCs,
atoll sources and the millisecond pulsar SAX J1808.4-3569.  If we
estimate the frequency of the lower kilohertz QPO by subtracting the
average frequency difference from the upper kilohertz QPO we can plot
this vs. the centroid frequency of this Lorentzian. We fit this
correlation and find that to within the errors it is consistent with
the correlation that Psaltis, Belloni \& van der Klis (1999) found for
neutron stars and black holes.

\begin{figure*}
\figurenum{7}
% need rescaling in preprint mode
\epsscale{1.5}
\plotone{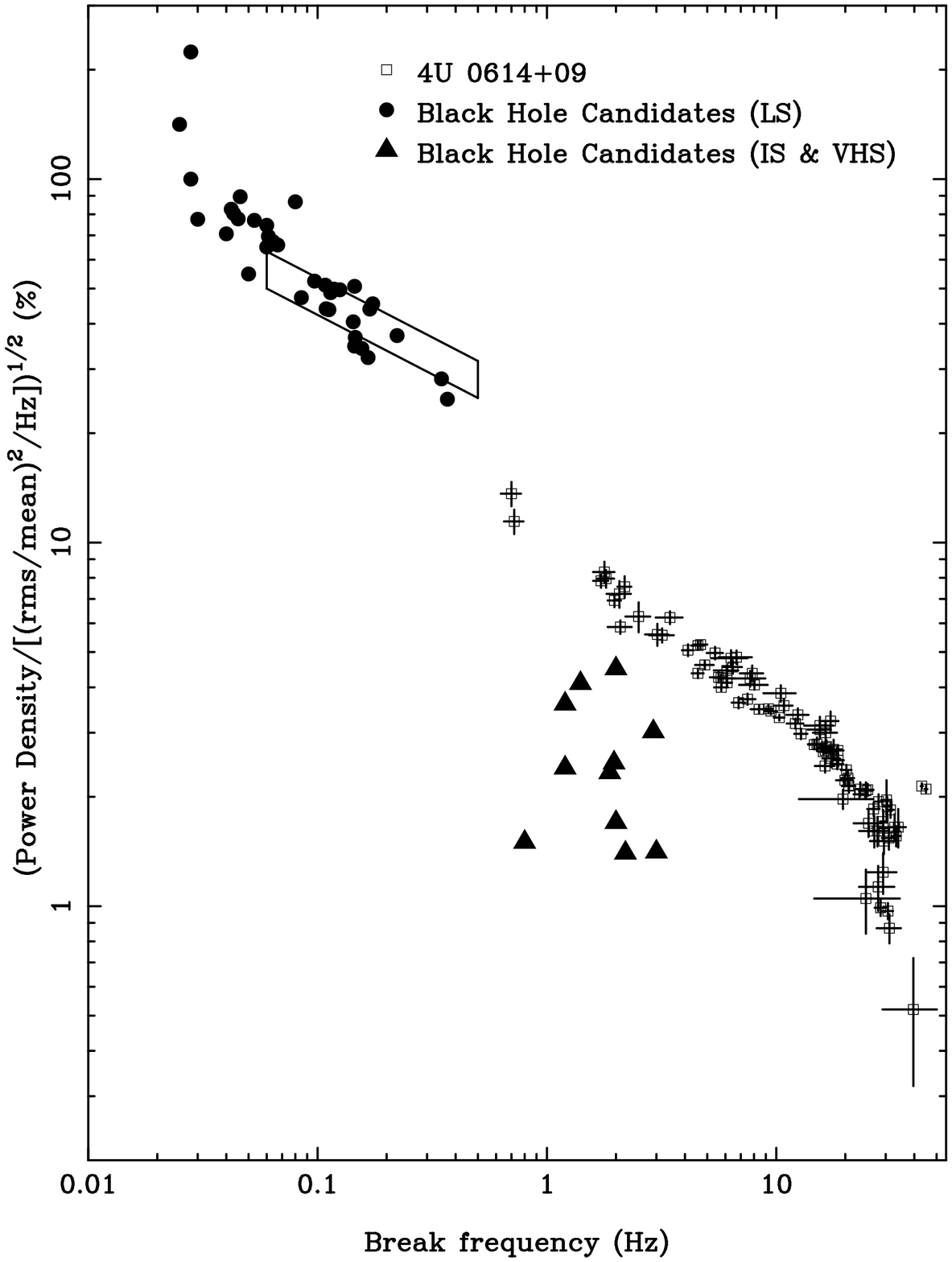} 
\caption{Power density of the broken power--law component at the
break vs. frequency of the break. The filled points and the marked region are from 
five different black hole candidates (\mm \& van der Klis 1997 and references therein). The filled 
circles and the marked region are observations in the low state and the filled triangles are from the 
intermediate and very high states. The open squares are the points from 4U 0614+09.}
\label{fig:bhpl}
\end{figure*}

In order to compare the HFN component with that seen in BHCs (\mm \&
van der Klis 1997), we also calculated the power density at the break
of the broken power--law. The values from 4U 0614+09 and from the BHCs
are plotted in Figure \ref{fig:bhpl}.

\section{Discussion}

\subsection{Timing--Spectra Correlations and States}

The previous section demonstrates connections between properties of
the energy spectrum and features in the power spectra of the low mass
X-ray binary 4U 0614+09. There is a strong correlation between the
position in the X-ray hard color vs. intensity diagram, the
frequencies of the kilohertz QPOs, the frequency of a 6$-$38 Hz
Lorentzian, the break frequency of the HFN and the strength of both
the HFN and the VLFN.

Historically the atoll sources (4U 0614+09 among them) are known to
show two source states, the so-called `banana' and `island' states
named after their shapes in the color-color diagram (Hasinger \& van
der Klis 1989). In the banana state there is a strong VLFN component
present in the power spectra at low frequencies and the energy
spectrum is soft. In the island state the power spectrum is dominated
by the HFN and the energy spectrum is harder. In the present data we
observe a smooth transition between the two states, indicating that
these are part of a continuous sequence (Hasinger \& van der Klis
1989). Similar transitions are seen in RXTE data of other atoll
sources (4U 1636$-$53, Prins \& van der Klis 1997; 4U 1705$-$44, Ford,
van der Klis \& Kaaret 1998; 4U 1820$-$30, Zhang et al. 1998a; 4U
1735$-$44, Ford et al. 1998; 4U 1608$-$52, \mm et al. 1999; 4U
1728$-$34, \mm \& van der Klis 1999; Aql X$-$1, Reig et al. 1999; \mm
1999).  In 4U 0614+09 the kilohertz QPOs are present in the whole
range of the X-ray color--color diagram in contrast to data sets on
similar sources that indicate that the QPOs are absent at the highest
$S_{\rm a}$ and possibly the lowest as well (4U 1608$-$52, \mm et
al. 1999; 4U 1705$-$44, Ford, van der Klis \& Kaaret 1998; 4U
1636$-$53, Zhang et al. 1996, Wijnands et al.  1997c; 4U 1735$-$44,
Wijnands et al. 1998, Ford et al. 1998; 4U 1820$-$30, Smale, Zhang \&
White 1997; KS 1731$-$260, Wijnands \& van der Klis 1997d). At the
highest $S_{\rm a}$ ($\geq$ 2.49) QPOs are present in 4U 0614+09 with
rms fractions of 4 to 8 \% in the total energy range (5 to 11 \% for
an energy range of 4.6$-$97.3 keV). These detections are similar to
previously reported upper limits for the rms fraction over the full
energy range in the banana state: $< 2-3 \%$ (4U 1735$-$44, Wijnands
et al. 1998b), $< 2-6 \%$ (4U 1705$-$44; Ford, van der Klis \& Kaaret
1998) and $< 1-5 \%$ (4U 1608$-$52, \mm et al. 1999). In 4U 0614+09 we
are probably not seeing the upper banana state, where there is no HFN
and the VLFN becomes very strong.

The correlation between the kilohertz QPO frequencies and $S_{\rm a}$
in this source is similar to that in the Z sources with $S_{\rm z}$
(van der Klis et al. 1996; Wijnands et al. 1997b; Jonker et
al. 1998). The strength and frequency of the signals at lower
frequency also show similar correlations.  $S_{\rm a}$ is a much
better indicator for the QPO frequencies than count rate, where a
correlation with the QPO frequencies on long time scales is lacking
(4U 0614+09, Ford et al. 1997a; Aql X$-$1, Zhang et al. 1998b; 4U
1820$-$30, Zhang et al. 1998a, Kaaret et al. 1999; 4U 1608$-$52, \mm
et al. 1999; and other sources, see \mm 1999).  Models explaining the
kilohertz QPOs predict, for different reasons, that the centroid
frequencies increase with mass accretion rate (Miller, Lamb \& Psaltis
1998; Titarchuk, Lapidus \& Muslinov 1998; Stella \& Vietri 1998).

The present data indicate a link between $S_{\rm a}$ and mass
accretion rate. The correlation between break frequency of the HFN and
$S_{\rm a}$ further strengthens this link. The suggestion that the
break frequency tracks the mass accretion rate was previously noted
for atoll sources (van der Klis 1994a; Prins et al. 1997). The only
possible exceptions to a correlation between break frequency and
inferred mass accretion rate so far are SAX J1808.4$-$3658 (Wijnands
\& van der Klis 1998) and SLX 1735$-$269 (Wijnands \& van der Klis
1999b) where at higher count rates the break frequency is smaller.
The exception to the correlation in these sources may be related to
the fact that they were observed at especially low inferred mass
accretion rates. The two spectra of SLX 1735$-$269 have exceptionally
low break frequencies of 2.3 and 0.11 Hz (Wijnands \& van der Klis
1999b). We note however, that accretion rates are inferred only from
the fluxes and the relative change in flux is small, about 20 \% in
SLX 1735$-$269 and about 40 \% in SAX J1808.4$-$3658. The
anti--correlation in these sources may be due to this small dynamic
range.

The changes in energy spectra, reflected in a changing $S_{\rm a}$,
are likely driven by changes in the mass accretion rate, as previously
suggested for 4U 0614+09 by the correlation between the frequency of
the kilohertz QPOs with the blackbody flux (Ford et al. 1997b) or with
the index of the power--law (Kaaret et al. 1998). Larger values of
$S_{\rm a}$ correspond to softer spectra, higher fluxes, and larger
inferred mass accretion rates. From color--color diagrams, at low
inferred mass accretion rate there is an anti-correlation of flux and
hardness, which is manifest also in spectral studies of atoll sources:
e.g., 4U 1636$-$53 (Breedon et al. 1986), 4U 1735$-$44 (Smale et
al. 1986), 4U 1705$-$44 (Langmeier et al. 1987), 4U 1608$-$52 (Mitsuda
et al.  1989), 4U 0614+09 (Barret \& Grindlay 1995; Ford et al. 1996),
and others (van der Klis \& van Paradijs 1994). One long-standing
explanation of this anti-correlation is in terms of thermal
Comptonization models (e.g. Sunyaev \& Titarchuk 1980). In the thermal
model the increasing soft X-ray flux cools the Comptonizing electron
plasma, leading to a decrease in its temperature and thus a decrease
in hard X-ray flux.

\subsection{Comparison with Other Sources}

The VLFN component might correspond to motion of the source along the
track in the color-color diagram during an observation, as suggested
for Z sources (van der Klis 1991).  Observations of Cyg X$-$2 however
indicate that the motion in the Z track is at most partly responsible
for the VLFN (Wijnands et al. 1997a).  Bildsten (1995) suggested that
the VLFN is caused by unstable burning of nuclear fuel in a regime of
time dependent helium burning. The dependence of the strength of the
VLFN on the mass accretion could then be explained by a slower mode of
combustion that sets in at higher mass accretion rate. This slower
mode of combustion leads to stronger long time scale variations.

The 6$-$38 Hz Lorentzians are similar to the horizontal branch
oscillations (HBOs) in the Z sources in their rms fractions and
correlations with $S_{\rm a}$ and $\nu_{\rm kHz}$. The HBOs sometimes
show a saturation of the HBO frequency with respect to $S_{\rm z}$ at
a frequency of about 60 Hz (GX 17+2, Wijnands et al. 1997b; Cyg X-2,
Wijnands et al. 1998a). In 4U 0614+09 the Lorentzian frequency only
goes up to 38 Hz and we do not observe a saturation of the Lorentzian
frequency with respect to $S_{\rm a}$ (Fig. \ref{fig:saqpo}), but we
do observe a saturation of the break frequency of the HFN with respect
to $S_{\rm a}$ (Fig. \ref{fig:salf}).

The 97$-$158 Hz Lorentzians could be related to the 65$-$317 Hz
oscillations seen in some of the black hole candidates (GRO J1655-40,
Remillard et al. 1999a; XTE J1550-564, Remillard et al. 1999b; GRS
1915+105, Morgan et al. 1997). These oscillations fall in the same
frequency range as the 97$-$158 Hz Lorentzians we observe but have
weaker rms fractional amplitudes ($\sim$1 \%). In GRS 1915+105 (65--68
Hz; Morgan et al. 1997) and GRO J1655-40 (281--317 Hz; Remillard et
al. 1999a) the oscillations were about constant in frequency. In XTE
J1550-564 however, the frequency varied considerably from 184 to 284
Hz (Homan et al. 1999; Remillard et al. 1999a).

For what concerns the other features in the power spectra, the high
frequency noise component in atoll sources is very similar to the
low-state noise in black hole candidates (van der Klis 1994a,b).  The
BHCs show a correlation of power density at the break with break
frequency (Belloni \& Hasinger 1989).  The relation between power
density at the break and break frequency is similar in 4U 0614+09 and
the BHCs in the low state (Fig. \ref{fig:bhpl}). However in the region
where the break frequencies of 4U 0614+09 and the BHCs become
comparable (the intermediate and very high state of the BHCs) 4U
0614+09 does show a larger power density. The HFN may result from a
superposition of shots with the break corresponding to the shots with
the longest durations, as proposed for the BHCs (e.g. Belloni \&
Hasinger 1990). The time scale of the shots could be set by the
lifetime of clumps (van der Klis 1994b) and could decrease with mass
accretion rate (Ford \& van der Klis 1998).

\subsection{Correlations of Timing Features}

Stella \& Vietri (1998) have proposed that the 6$-$38 Hz Lorentzian is
caused by relativistic frame dragging in the inner accretion disk
(Lense-Thirring precession). The frequency of the Lorentzian depends
on the frequency of the upper kilohertz QPO, assumed to be fixed by
the Keplerian frequency at the inner edge of the disk. Stella \&
Vietri (1998) predict a scaling index of 2, but see corrections by
Morsink \& Stella (1998). Our value of 2.46 $\pm$ 0.07 is
significantly different from 2. Also, as in 4U 1728$-$34 (Ford \& van
der Klis 1998), the required ratio of the moment of interia over the
neutron star mass is unphysically large if the nodal precession
frequency is the QPO frequency. However this is not a problem if the
QPO frequency is twice the nodal precession frequency (Stella \&
Vietri 1998).

Titarchuk \& Osherovich (1999) associate the 6$-$38 Hz Lorentzian with
radial oscillations in a boundary layer where the disk adjusts to the
rotating neutron star and the break frequency of the HFN with the
diffusion time scale in this region. The correlations in 4U 0614+09 of
the {\em lower} kilohertz QPO vs. 6$-$38 Hz Lorentzian and HFN break
frequency do not fall on the predicted curves. In fact the
correlations with respect to the {\em higher} frequency kilohertz QPO
are closer to those predicted.  In addition there is a high frequency
turnover of the break frequency which is not predicted
(Fig. \ref{fig:khzvsqpo}).

\subsection{The Highest Frequency QPOs}

We report here the highest frequency QPO of any low mass X-ray binary
to date at 1329 $\pm$ 4 Hz.  Such a high frequency oscillation puts
new constraints on the structure of neutron stars (Miller, Lamb \&
Cook 1998). The mass and radius are bounded by:
$$ M \leq 2.2 (\nu_{K}/1000\,{\rm Hz})^{-1} (1+0.75j) 
 {\rm ~} M_{\odot}$$
and
$$ R \leq 19.5 (\nu_{K}/1000\,{\rm Hz})^{-1} (1+0.2j) {\rm ~km} $$
where $j$ is the dimensionless angular momentum of the star, $j \equiv
2\pi cI\nu_{spin}/GM^{2}$ (e.g.  Miller, Lamb \& Psaltis 1998). This
expansion in $j$ is valid for spin frequencies less than about 400 Hz
(Miller, Lamb \& Cook 1998). For $I =$ 3$\times 10^{45}$ g cm$^2$
(Cook, Shapiro \& Teukolsky 1994; Miller, Lamb \& Psaltis 1998) and
$\nu_{spin}=311.8$ Hz, the inferred spin frequency of the star, the
kilohertz QPO at 1329 $\pm$ 4 Hz constrains $M$ and $R$ to $M \leq 1.9
M_{\odot}$ and $R \leq 15.2$~km. These constraints rule out the
extremely stiff EOS. A mean-field theory (`L' in Cook, Shapiro \&
Teukolsky 1994) is ruled out and the tensor interaction EOS (`M' in
Cook, Shapiro \& Teukolsky 1994) is ruled out except for a very small
range of R and M.  However we note, that this QPO has a single-trial
significance of only 3.5 $\sigma$.  To put constraints on the softer
EOS the kilohertz QPO must be observed at higher frequencies. For a
slow rotating star, 2010 Hz would rule out the `U'-type EOS (Wiringa,
Fiks \& Fabrocini 1988) and 1920 Hz would rule out the commonly used
`FPS' EOS (Cook, Shapiro \& Teukolsky 1994).

We note that there is no obvious saturation of the kilohertz QPO
frequency with respect to $S_{\rm a}$, our chosen indicator of mass
accretion rate (see Fig. \ref{fig:sakhzqpo}). However the kilohertz
QPO at 1329 Hz does deviate from the linear trend seen for the other
kilohertz QPOs. Such a saturation where the frequency becomes
independent of the chosen mass accretion rate indicator would be an
indication for the marginally stable orbit predicted by general
relativity (Kluzniak, Michelson \& Wagoner 1990; Kaaret, Ford \& Chen
1997; Miller, Lamb \& Psaltis 1998). Such a saturation has been
reported in 4U 1820$-$30 (Zhang et al. 1998a; Kaaret et al. 1999).

\subsection{QPO Frequency Difference}

\begin{figure*}
\figurenum{8}
% need rescaling in preprint mode
\epsscale{1.5}
\plotone{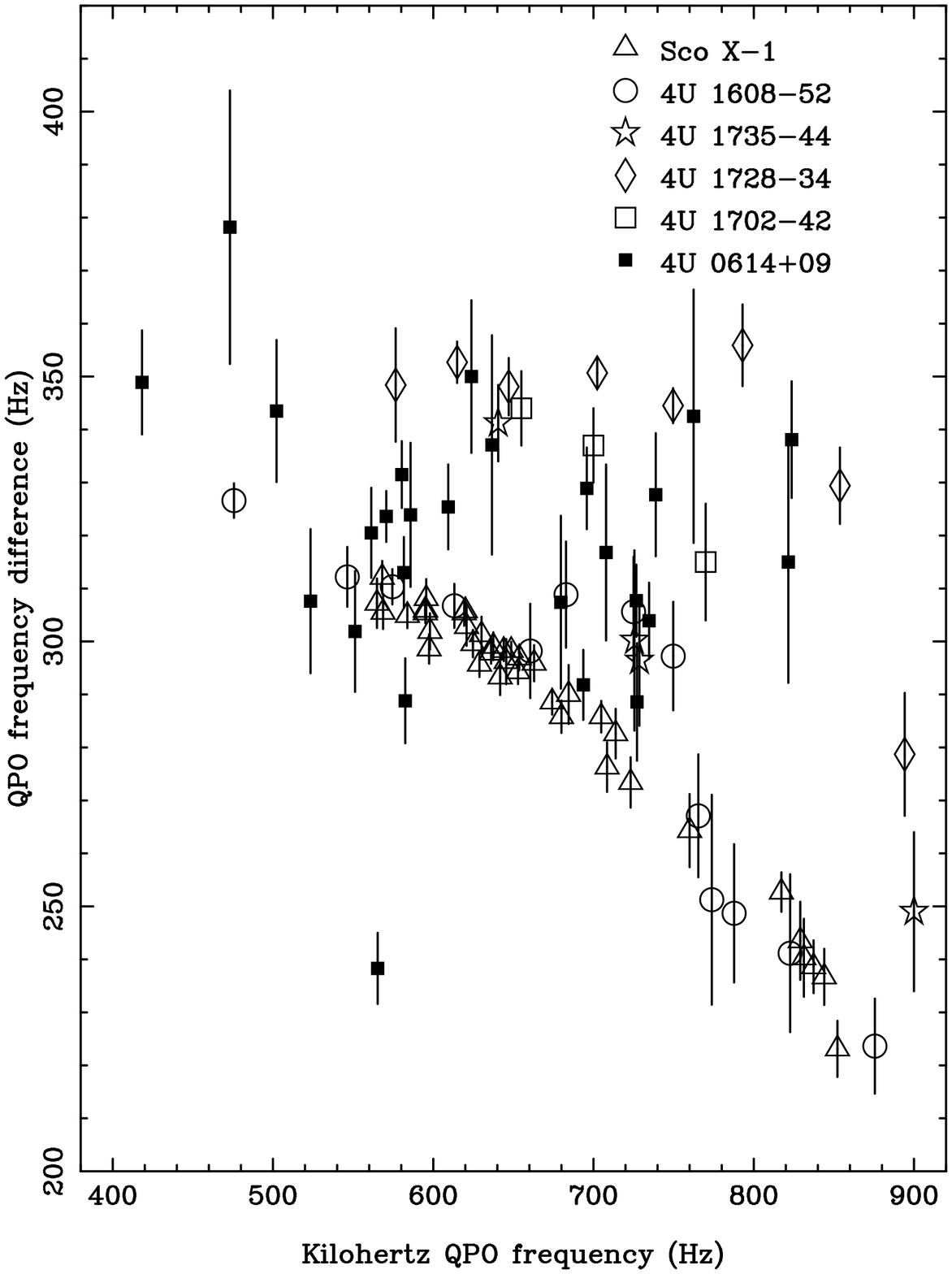} 
\caption{The difference in frequency between double kilohertz QPOs 
vs. the frequency of the lower kilohertz QPOs for four different sources. The open triangles are 
the points from Sco X--1 (van der Klis et al. 1997), the open circles are the points from 
4U 1608--52 (\mm et al. 1998), the open stars are the points from 4U 1735--44 (Ford et al. 1998; Wijnands 
et al. 1998b), 
the open diamonds are the points from 4U 1728--34 (\mm \& van der Klis 1999), the open squares are the 
points from 4U 1702--42 (Markwardt et al. 1999a,b) and the filled squares are the points from 4U 
0614+09 (this paper).}
\label{fig:freqdiff_alls}
\end{figure*}

We found that, $\Delta \nu$, the frequency difference between the
upper and the lower kilohertz QPO peak is constant in 4U 0614+09,
confirming the results of Ford et al. (1997a). This matches the
prediction of the beat frequency mechanism (e.g. Miller, Lamb \&
Psaltis 1998) in which $\Delta \nu$ is equal to the spin frequency and
therefore constant. However observations of five sources have shown a
significant decrease in $\Delta \nu$ as the frequency of the kilohertz
QPOs increases (Sco X--1, van der Klis 1997; 4U 1608-52, \mm et
al. 1998a,b; 4U 1735-44, Ford et al. 1998; 4U 1728-34, \mm \& van der
Klis 1999; 4U 1702-42, Markwardt et al. 1999a,b). To examine the trend
of $\Delta \nu$ with increasing QPO frequency in 4U 0614+09 we plot
$\Delta \nu$ vs. the frequency of the lower frequency kilohertz QPO
for all these sources in Figure \ref{fig:freqdiff_alls}.  We find that
at low QPO frequency ($< 750$ Hz) $\Delta \nu$ in 4U 0614+09 is
consistent with Sco X--1 and 4U 1608--52 and at high QPO frequency
(750--820 Hz) inconsistent (2.8--8.0 $\sigma$ different). However, we
note that one of the points from 4U 0614+09 itself deviates 11.0
$\sigma$ from the trend seen in 4U 0614+09.

\section{Acknowledgements}

This work was supported by NWO SPINOZA grant 08--0 to E.P.J. van den
Heuvel, by the Netherlands Organization for Scientific Research (NWO)
under contract number 614--51--002, and by the Netherlands Research
School for Astronomy (NOVA). MM is a fellow of the Consejo Nacional de
Investigaciones Cient\'{\i}ficas y T\'ecnicas de la Rep\'ublica
Argentina.  PK acknowledges support from NASA grants NAG5-7405,
NAG5-7334, and NAG5-7477.  This research has made use of data obtained
through the High Energy Astrophysics Science Archive Research Center
Online Service, provided by the NASA/Goddard Space Flight Center.

%%%%%%%%  TABLE 1 and 2: Observations  %%%%%%%%%%%

\begin{figure*}
\figurenum{-n}
% need rescaling in preprint mode
\epsscale{2.0}
\plotone{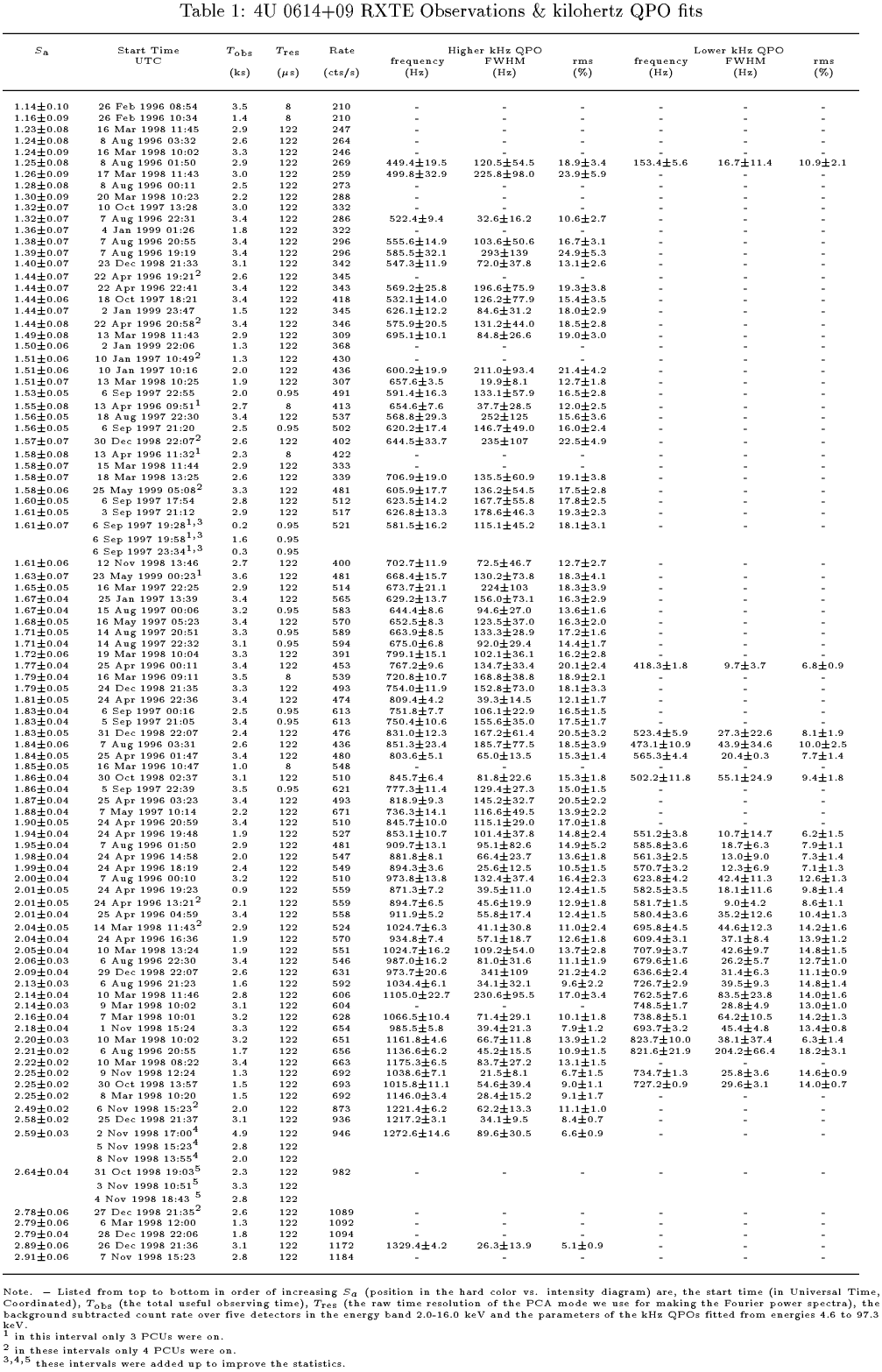} 
\caption{ }
\label{tbl:obsa}
\end{figure*}

\begin{figure*}
\figurenum{-n}
% need rescaling in preprint mode
\epsscale{2.0}
\plotone{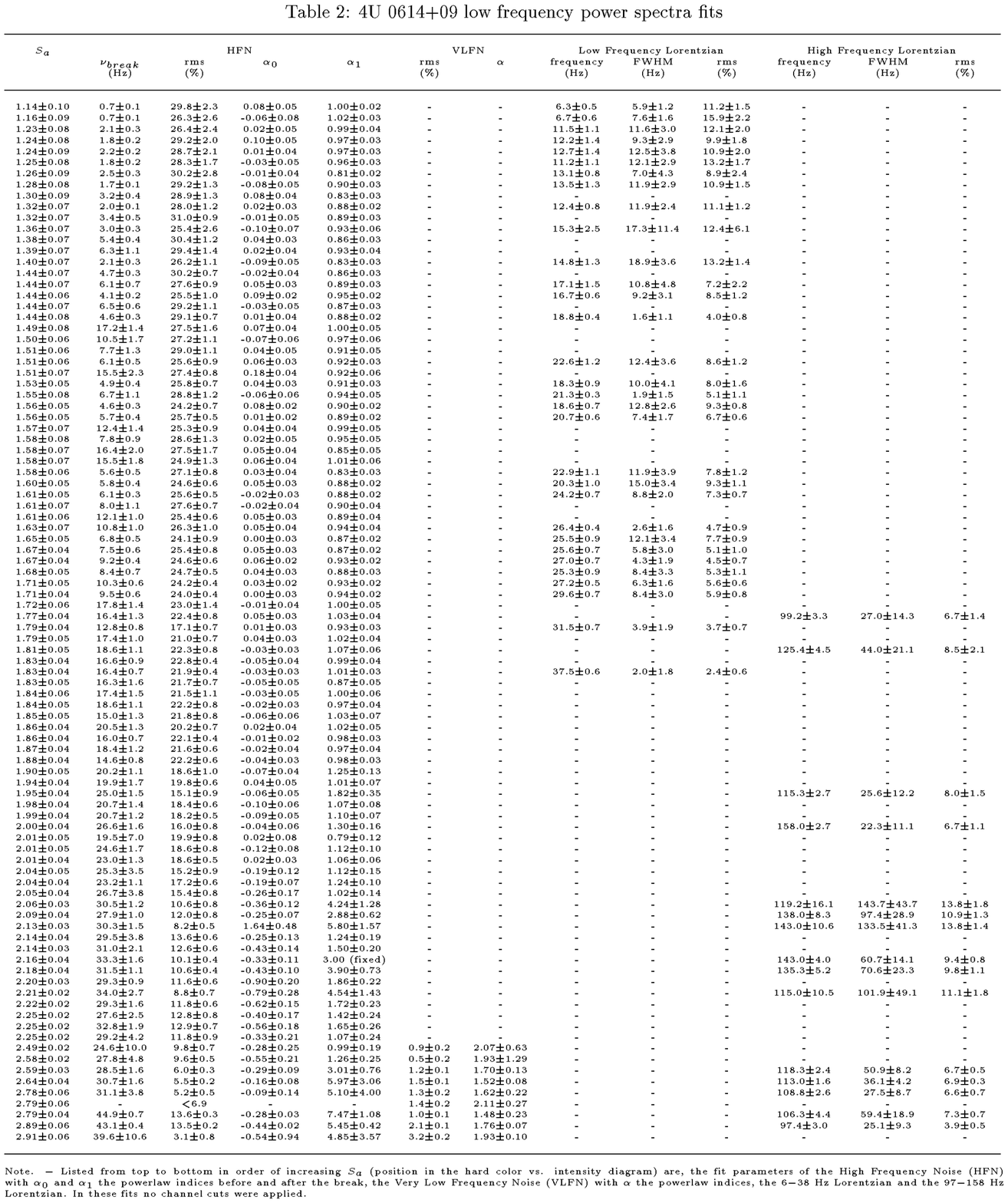} 
\caption{ }
\label{tbl:obsb}
\end{figure*}

%%%%%%%%  FIGURES:

\begin{references}

\reference{} 
\reference{barret95} Barret, D., \& Grindlay, J.E., ApJ, 440, 841 (1995)
\reference{belloni90} Belloni, T., \& Hasinger, G., A\&A, 227, L33 (1990)
\reference{bidsten95} Bildsten, L., ApJ, 438, 852 (1995)
\reference{brandt92} Brandt, S., Castro-Tirado, A. J., Lund, N., Dremin, V., Lapshov, I., \& 
Sunyaev, R., A\&A, 262, L15 (1992)
\reference{brandt94} Brandt, S., Ph.D thesis, Danish Space Res. Inst. (1994)
\reference{breedon86} Breedon, L.M., Turner, M.J.L., 
King, A.R., Courvoisier, T.J.L., MNRAS, 218, 487 (1986)
\reference{cook94} Cook, G.B, Shapiro, S.L., \& Teukolsky, S.A., ApJ, 424, 823 (1994) 
\reference{ford96} Ford, E.C., et al., ApJ, 469, L37 (1996)
\reference{ford97a} Ford, E.C., et al., ApJ, 475, 123 (1997a)
\reference{ford97b} Ford, E.C., et al., ApJ, 486, 47 (1997b)
\reference{ford98} Ford, E.C., van der Klis, M., \& Kaaret, P., 
ApJL, 498, L41 (1998) 
\reference{ford98} Ford, E.C., \& van der Klis, M., ApJ, 506, 39 (1998)
\reference{ford98} Ford, E.C., van der Klis, M., van Paradijs, J., \mm, M., Wijnands, R.,
Kaaret, P., ApJ, 508, L155 (1998)
\reference{grove94} Grove, J., et al., AIP Conference Proc. 304, p192 (1994)
\reference{hasinger89} Hasinger, G., \& van der Klis, M., A\&A, 225, 79 (1989)
\reference{hasinger90} Hasinger, G., van der Klis, M.,
Ebisawa, K., Dotani, T., Mitsuda, K., A\&A, 235, 131 (1990)
\reference{hertz92} Hertz, P., Vaughan, B., Wood, K.S.,
Norris, J.P., Mitsuda, K., Michelson, P.F., Dotani, T., ApJ, 396, 201 (1992) 
\reference{homan99} Homan, J., Wijnands, R., van der Klis, M., IAU Circ. No. 7121 (1999)
\reference{jonker98} Jonker, P.G., Wijnands, R.,
van der Klis, M., Psaltis, D., Kuulkers, E., Lamb, F.K., ApJ, 499, 191 (1998)
\reference{kaaret97} Kaaret, P., Ford, E., \& Chen, K., ApJL, 480, L27 (1997)
\reference{kaaret98} Kaaret, P., Yu, W., Ford, E.C., Zhang, S.N., ApJL, 497, L93 (1998)
\reference{kaaret99} Kaaret, P., Piraino, S., Bloser, P.F., Ford, E.C., Grindlay, J.E.,
Santangelo, A., Smale, A.P., Zhang, W., ApJL, 520, 37 (1999)
\reference{kluzniak90} Kluzniak, W., Michelson, P., \& Wagoner, R.V., ApJ, 358, 538 (1990)
\reference{kuulkers94} Kuulkers, E., van der Klis, M., Oosterbroek, T.,
 Asai, K., Dotani, T., van Paradijs, J., Lewin, W.H.G., A\&A, 289, 795 (1994)
\reference{kuulkers96} Kuulkers, E., \& van der Klis, M, A\&A, 314, 567 (1996)
\reference{kuulkers97} Kuulkers, E., van der Klis, M., Oosterbroek, T., van Paradijs, J., 
Lewin, W.H.G., MNRAS, 287, 495 (1997)
\reference{langmeier87} Langmeier, A., Sztajano, M., Hasinger, G., Truemper, J., 
Gottwald, M., ApJ, 323, 288 (1987)
\reference{leahy83} Leahy, D., Darbro, W., Elsner, R.F., Weisskopf, M.C., Kahn, S., 
Sutherland, P.G., Grindlay, J.E., ApJ, 266, 160 (1983)
\reference{markwardt99a} Markwardt, C.B., Strohmayer, T.E., Swank J.H., ApJ, 512, 125 (1999a)
\reference{markwardt99b} Markwardt, C.B., Lee, H.C., wank J.H., AAS HEAD, 31:15.01 (Abstr.) (1999b)
\reference{mendez97} \mm, M., \& van der Klis, M., ApJ, 479, 926 (1997)
\reference{mendez97} \mm, M., van der Klis, M., van Paradijs, J., Lewin, W.H.G., 
Lamb, F.K., Vaughan, B.A., Kuulkers, E., Psaltis, D., ApJ, 485, L37 (1997)
\reference{mendez98a} \mm, M., et al.  ApJ, 494, 65  (1998a)
\reference{mendez98b} \mm, M., van der Klis, M., Wijnands, R.A.D., Ford, E.C., van Paradijs, J.
, Vaughan, B.A., ApJ, 505, 23 (1998b)
\reference{mendez99} \mm, M., van der Klis, M., Ford, E.C., 
Wijnands, R., van Paradijs, J., ApJ, 511, 49 (1999)
\reference{mendez99} \mm, M., \& van der Klis, M., ApJ, 517 , 51 (1999)
\reference{mendez99} \mm, M., astro-ph/9903469, to appear in 
Proceedings of the 19th Texas Symposium in Paris (1999) 
\reference{miller98} Miller, M.C., Lamb, F., and Psaltis, D., ApJ, 508, 791 (1998)
\reference{miller98b} Miller, M.C., Lamb, F., and Cook, G., ApJ, 509, 793 (1998)
\reference{mitsuda89} Mitsuda, K., Inoue, H., Nakamura, N., Tanaka, Y., PASJ, 41, 97 (1989)
\reference{morgan97} Morgan, E.H.; Remillard, R.A.; Greiner, J., ApJ, 482, 993 (1997)
\reference{olive98} Olive, J.F., Barret, D., Boirin, L., Grindlay, J.E., Swank, J.H., 
Smale, A.P., A\&A, 333, 942 (1998)
\reference{prins97} Prins, S., \& van der Klis, M., A\&A, 319, 498 (1997)
\reference{psaltis99} Psaltis, D., Belloni, T., \& van der Klis, M., ApJ, 520, 262 (1999)
\reference{reig99} Reig, P., et al., in preparation (1999)
\reference{remillard99a} Remillard, R.A., Morgan, E.H., McClintock, J.E., Bailyn, C.D., 
Orosz, J.A., ApJ, 522, 397 (1999a)
\reference{remillard99b} Remillard, R.A., McClintock, J.E., Sobczak, G.J., Bailyn, C.D., 
Orosz, J.A., Morgan, E.H., Levine, A.M., ApJ, 517, 127 (1999b)
\reference{singh94} Singh, K.P., \& Apparao, K.M.V., ApJ, 431, 826 (1994)  
\reference{smale86} Smale, A.P., Corbet, R.H.D., Charles, P.A., Menzies, J.W., 
Mack, P, MNRAS, 223, 207 (1986)
\reference{smale97} Smale, A.P., Zhang, W. and White, N.E., ApJL, 483, L119 (1997)
\reference{swank78} Swank, J. H., Becker, R. H., Boldt, E. A., Holt, S. S., 
\& Serlemitsos, P. J., MNRAS, 182, 349 (1978)
\reference{swank98} Swank, J., astro-ph/9802188, to appear in 
"The Active X-Ray Sky", eds. L. Scarsi, H. Bradt, P. Giommi, and
F. Fiore (1998)
\reference{stella98} Stella, L., \& Vietri, M., ApJL, 492, L59 (1998)
\reference{sunyaev80} Sunyaev, R. and Titarchuk, L., A\&A, 86, 121 (1980)
\reference{titarchuk98} Titarchuk, L., Lapidus, I., \& Muslimov, A., ApJ, 499, 
315 (1998)
\reference{titarchuk99} Titarchuk, L., \& Osherovich, V., ApJL, 518, 95 (1999) 
\reference{vdk85} van der Klis, M., Nature, 316, 225 (1985)
\reference{vdk89} van der Klis, M., in Timing Neutron Stars, Kluwer Academic Publishers: Boston (1989)
\reference{vdk91} van der Klis, M., In: "Neutron Stars: Theory and Observations", 
eds. J. Ventura \& D. Pines, NATO ASI Series C, 344, p. 319 (1991)
\reference{vdk94a} van der Klis, M., A\&A, 283, 469 (1994a)
\reference{vdk94b} van der Klis, M., ApJS, 92, 511 (1994b)
\reference{vdk94} van der Klis, M., \& van Paradijs, J., A\&A, 281, L17 (1994)
\reference{vdk96a} van der Klis, M., Swank, J.H., Zhang, W., Jahoda, K., 
Morgan, E.H., Lewin, W.H.G., Vaughan, B., van Paradijs, J., ApJL, 469, 1 (1996)
\reference{vdk97} van der Klis, M., Wijnands, R.A.D., Horne K., Chen W., ApJ, 481, 97, (1997) 
\reference{vdk98} van der Klis, M., in The Many Faces of Neutron Stars, ed. R. Buccheri, 
J. van Paradijs, \& M. A. Alpar (NATO ASI Series C, Vol. 515, Dordrecht: Kluwer), 337 (1998)
\reference{vdk99} van der Klis, M.,  in Proc. of the Third William Fairbank Meeting, 
Rome June 29 - July 4 1998 (1999)
\reference{wijnands97a} Wijnands, R.A.D., van der Klis, M., Kuulkers, E., 
Asai, K., Hasinger, G., A\&A, 323, 399 (1997a)
\reference{wijnands97b} Wijnands, R.A.D., et al., ApJ, 490, L157 (1997b)
\reference{wijnands97c} Wijnands, R.A.D., van der Klis, M., van Paradijs, J., Lewin, W.H.G., 
Lamb, F.K., Vaughan, B., Kuulkers, E., ApJL, 479, L141 (1997c)
\reference{wijnands97} Wijnands, R.A.D., \& van der Klis, M., ApJL, 482, L65 (1997)
\reference{wijnands98} Wijnands, R.A.D., \& van der Klis, M., ApJ, 507, L63 (1998)
\reference{wijnands98a} Wijnands, R.A.D., et al., ApJL, 493, L87 (1998a) 
\reference{wijnands98b} Wijnands, R.A.D., van der Klis, M., \mm, M., van Paradijs, J., 
Lewin, W.H.G., Lamb, F.K., Vaughan, B., Kuulkers, E., ApJL, 495, L39 (1998b) 
\reference{wijnands99a} Wijnands, R.A.D., \& van der Klis, M., ApJ, 514, 939 (1999a)
\reference{wijnands99b} Wijnands, R.A.D., \& van der Klis, M., astro-ph/9903450 (1999b)
\reference{wiringa88} Wiringa, R.B., Fiks, V., \& Fabrocini, A., Phys. Rev. C, 38, 1010 (1988)
\reference{yoshida93} Yoshida, K., Mitsuda, K., Ebisawa, K., Ueda, Y., Fujimodo, R., Yaqoob, T., 
Done, C., PASJ, 45, 605 (1993)
\reference{zhang93} Zhang, W., Giles, A.B., Jahoda, K., Soong, Y., Swank, J.H., 
Morgan, E.H., SPIE, 2006, 324 (1993)
\reference{zhang96} Zhang, W., Lapidus, I., White, N.E., \& Titarchuk, L., ApJL, 469, L17 (1996) 
\reference{zhang98a} Zhang, W., Smale, A.P., Strohmayer, T.E., Swank, J.H., ApJ, 500, L171 (1998a)
\reference{zhang98b} Zhang, W., Jahoda, K., Kelley, R.L., Strohmayer, T.E., Swank, J.H. 
and Zhang, S.N., ApJ, 495, L9 (1998b)

\end{references}
\end{document}